\documentclass[11pt]{article}

\usepackage{epsfig}
\usepackage{latexsym}
\usepackage{amssymb}

\def\be{\begin{equation}}
\def\ee{\end{equation}}
\def\ba{\begin{eqnarray}}
\def\ea{\end{eqnarray}}

\title{{\bf Gravitating fluxbranes}}
\author{{\bf P. M. Saffin}\thanks{email: p.m.saffin@durham.ac.uk}
\\ Centre for Particle Theory, Department of Mathematical Sciences, \\
University of Durham, South Road, Durham DH1 3LE, 
United Kingdom.\\
\\ \\ Preprint DCPT/01/33}

\date{\today}

\begin{document}

\maketitle
\begin{abstract}
We consider the effect that gravity has when one tries to
set up a constant background form field. 
We find that in analogy with
the Melvin solution, where magnetic field lines self-gravitate
to form a flux-tube, the self-gravity of the form field
creates fluxbranes. Several exact solutions are found
corresponding to different transverse spaces and world-volumes,
a dilaton coupling is also considered.
\end{abstract}

\section{Introduction}

The Melvin solution \cite{melvin64} in Einstein-Maxwell theory
was constructed many years ago, describing what happens
to a uniform magnetic field when gravity is included.
Namely, the lines of magnetic flux self gravitate and form a
cylindrically symmetric configuration which is classically stable
\cite{thorn65}.
Over the years a number of generalizations have appeared, both
within the context of extending to metrics which are asymptotically
Melvin and changing the field content of the model \cite{general}
\cite{dowker95}. Of particular relevance is the work of Gibbons and
Maeda \cite{gibbons88} where the Melvin flux-tube solution was extended
to arbitrary dimensions, with the inclusion of a dilaton, to give
fluxbranes.
The Melvin solution has also found it's way into the string theory
literature, with it appearing as an exact background for
propagating strings, \cite{string}.

In these days of supergravity theories we find that there
is another natural generalization to be done.
The Maxwell two-form field strength is just one of a more 
general class
of $n$-form field strengths, so it is of interest to study
the effect gravity has on a homogeneous $n$-form background.

The solutions we find here for an F$_{(n)}$ form
have the character of branes
with a worldvolume dimension of $D-n$, called fluxbranes.
We find exact solutions for these fluxbranes
with both curved and flat worldvolume,
leaving the
spacetime dimension and the rank of F$_{(n)}$ general. The inclusion
of a dilaton has also been studied.
In distinction to the Melvin solution these fluxbranes are 
singular at their core, asymptotically however the solutions
are regular.

To get a picture of what these fluxbranes are consider the case of Melvin,
where we have an infinite flux-tube. We can think of this heuristically as coming
from a monopole-antimonopole pair with infinite separation, and the flux
joins them in the form of a tube. In the more general case we could think
of the situation where we have a brane-antibrane pair of infinite separation,
where these branes are charged under an $n$ form field strength. For example
in 11D SUGRA  we have an 4-form field strength giving five-branes. 
Separating a $5\;\overline{5}$ pair with infinite
distance gives a \mbox{$D-n-1=6$} fluxbrane joining them, i.e. a fluxbrane
with $D-n=7$ world-volume dimensions.

\setcounter{equation}{0}
\section{The Melvin solution and its generalization}

In four dimensions the Melvin solution corresponds to a cylindrically
symmetric flux tube, which we take to be in the $z$-direction. The
metric and field strength take the following form,
\begin{eqnarray}
\label{melvinmetric}
{\rm ds}^2_{\rm Melvin} &=&(1+\rho^2)^2\left[-{\rm dt}^2+{\rm dz}^2\right]
                       +(1+\rho^2)^2{\rm dr}^2+\frac{r^2}{(1+\rho^2)^2}{\rm d}\phi^2,\\
\label{melvinform}
{\rm F}&=&\frac{1}{(1+\rho^2)^2}{\rm d}r\wedge r{\rm d}\phi.
\end{eqnarray}
The core has world volume coordinates $(t,z)$ and the form F lives on
the angular coordinate $\phi$ and the radial coordinate $r$ ($\rho\propto Br$),
so the magnetic field (of strength $B$) 
points in the $z$-direction. A detailed description
of the structure of this spacetime can be found in \cite{thorn65}.

Taking the Melvin case as a template we
extend this to consider an $(m+1)$ form 
field strength in general spacetime dimensions.
In the Melvin case the form lives on a circle and a radial
direction, here we take the form to have components on a compact
Euclidean manifold (with volume form $\eta_{(m)}$) and a radial direction
(with coordinate $\xi$),
\begin{eqnarray}
{\rm F}_{\rm mag'\;flux}&=&\alpha(\xi){\rm d}\xi\wedge \eta_{(m)}.
\end{eqnarray}
Just as in the Melvin solution the Hodge dual of the form
has components only in the worldvolume coordinates.
There is also a dual solution where the fluxbrane carries electric
rather than magnetic flux.
The duality relates a magnetic flux tube for
F$_{(m+1)}$ to an electric flux tube for a $D-(m+1)$ form field strength,
G$_{(D-(m+1))}$,
\begin{eqnarray}
{\rm *G}_{\rm elec'\;flux}&=&\alpha(\xi){\rm d}\xi\wedge \eta_{(m)}.
\end{eqnarray}
This compares to the usual case of branes
charged magnetically under an $m+1$ form field strength,
\mbox{${\rm F}_{\rm mag'\;charge}=\beta(r)\eta_{(m+1)}$}
where the world volume has $D-(m+2)$ dimensions, a $D-(m+3)$ brane.
The metric we take has a cylindrical symmetry,
\begin{eqnarray}
\label{basic_metric}
{\rm ds}^2 &=& \exp(2a(\xi))\overline{{\rm ds}}^2_{\rm (L)}
              +\exp(2b(\xi)){\rm d}\xi^2
              +\exp(2c(\xi))\overline{{\rm ds}}^2_{\rm (E)}
\end{eqnarray}
where $\overline{{\rm ds}}^2_{\rm (L)}$ is an $l+1$ dimensional 
Einstein manifold of (mostly positive)
Lorentzian signature, $\overline{{\rm ds}}^2_{\rm (E)}$
is an $m$ dimensional Einstein manifold of Euclidean
signature.
There is still a coordinate freedom in the choice of
$\xi$,
this shall be removed later when
a suitable gauge choice will be made
to simplify our resulting equations. 

\setcounter{equation}{0}
\section{Equations of motion}
\subsection{Curvature equations}

In order to be self contained we include here the calculation
of the Riemann curvature two forms 
which are needed to get the equations
governing the fluxbranes.
We use the convention that the zero index is time-like and
the curvatures of (\ref{basic_metric}) are normalised according to
\begin{eqnarray}
\overline{R}_{\rm (L)}^{ij}&=&l\Lambda_{\rm (L)}\overline{g}_{\rm (L)}^{ij},\\
\overline{R}_{\rm (E)}^{\alpha\beta}&=&
(m-1)\Lambda_{\rm (E)}\overline{g}_{\rm (E)}^{\alpha\beta},
\end{eqnarray}
where the over-line refers to the metrics $\overline{{\rm ds}}^2_{\rm (L),(E)}$.
Our notation for indices is that; $i,j=0...l$ and cover the Lorentzian
metric, $\alpha,\beta=l+2...D-1$ which cover the Euclidean metric, the
index $(l+1)$ is for the radial coordinate $\xi$. Capital Roman 
letters shall be used to denote general indices, $M,N=0...D-1$.

Our orthonormal basis one forms for ds$^2$
are denoted $e^i=\exp(a)\overline{e}^i$, 
$e^\xi=\exp(b){\rm d}\xi$ and $e^\alpha=\exp(c)\overline{e}^\alpha$,
where the $\overline{e}^i,\overline{e}^\alpha$ 
are understood to be the orthonormal bases
on $\overline{\rm ds}^2_{\rm (L),(E)}$.
In this basis then we find that the curvature two forms are
\begin{eqnarray}
\label{curvature_two_forms}
R^i_{\;j}&=&\overline{R}^i_{\;j}
              -(a')^2 \exp(-2b) e^i\wedge e^k\eta_{kj},\\
\nonumber
R^\alpha_{\;\beta}&=&\overline{R}^\alpha_{\;\beta}
                 -(c')^2 \exp(-2b) e^\alpha\wedge e^\gamma\eta_{\gamma\beta},\\
\nonumber
R^i_{\;\xi}&=&\left[a''-a'b'-(a')^2\right]\exp(-2b) e^\xi\wedge e^i,\\
\nonumber
R^\alpha_{\;\xi}&=&\left[c''-c'b'+(c')^2\right]\exp(-2b) e^\xi\wedge e^\alpha,\\
\nonumber
R^i_{\;\alpha}&=&-a'c'\exp(-2b) e^i\wedge e^\gamma\eta_{\gamma\alpha},
\end{eqnarray}
where $\eta_{\rm MN}$ is the usual Minkowski metric.
Using the fact that the $\overline{{\rm ds}}^2_{\rm (L),(E)}$ are Einstein 
manifolds we find the components of the Ricci tensor,
\begin{eqnarray}
{\cal R}_{ij}&=&-\left[a''-a'b'+(l+1)(a')^2+ma'c'\right]\exp(-2b)\eta_{ij}
                  +l\Lambda_{\rm (L)}\overline{g}_{{\rm (L)}ij}\exp(-2a),\\
{\cal R}_{\alpha\beta}&=&-\left[c''-c'b'+m(c')^2+(l+1)a'c'\right]\exp(-2b)\eta_{\alpha\beta}
                  +(m-1)\Lambda_{\rm (E)}\overline{g}_{{\rm (E)}\alpha\beta}\exp(-2c),\\
{\cal R}_{\xi\xi}&=&\left[-(l+1)a''-mc''+(l+1)a'b'+mb'c'-(l+1)(a')^2-m(c')^2\right]exp(-2b),
\end{eqnarray}
taking care to notice that the curvature forms
$\overline{R}$ are defined with an orthonormal
basis of $\overline{{\rm ds}}^2_{\rm (L),(E)}$ not ${\rm ds}^2$. So,
for example,
\begin{eqnarray}
\overline{R}^\alpha_{\;\beta}&=&\frac{1}{2}\overline{{\cal R}}^\alpha_{\;\beta\gamma\delta}
                              \overline{e}^\gamma\wedge \overline{e}^\delta\\
     &=&\exp(-2c)\frac{1}{2}\overline{{\cal R}}^\alpha_{\;\beta\gamma\delta}
                              e^\gamma\wedge e^\delta.
\end{eqnarray}

\subsection{Source equations}

By analogy with the Melvin solution (\ref{melvinform})
we require the form F to represent
a magnetic flux on the brane, that is to say the Hodge dual
of F is to have its components in the worldvolume directions.
We are led to consider an $(m+1)$ form ansatz,
\begin{eqnarray}
\label{fieldstrength}
F=f(\xi)e^\xi\wedge e^{l+2}\wedge e^{l+3}\wedge...\wedge e^{D-1},
\end{eqnarray}
which clearly satisfies the Bianchi equation because
$e^{l+2}\wedge e^{l+3}\wedge...\wedge e^{D-1}$ is,
up to a function of $\xi$, the
volume form on $\overline{{\rm ds}}^2_{\rm (E)}$.
The equation of motion ${\rm d*F}=0$ is satisfied for
$f(\xi)=\kappa \exp[-(l+1)a]$, where $\kappa$ is a constant, representing
the strength of the flux.
With the form equations satisfied we are left with the gravity equations
for an $m+1$ form field strength,
\begin{eqnarray}
{\cal R}_{MN}&=&\frac{1}{2(m!)}\left[F_{M...}F_N^{...}-\frac{m}{(m+1)(l+m)}F^2g_{MN}\right].
\end{eqnarray}
This gives two equations of motion and one constraint,
\begin{eqnarray}
a''-a'b'+(l+1)(a')^2+ma'c'-l\Lambda_{\rm (L)}\exp(2b-2a)
&=&\frac{1}{2}\frac{m}{l+m}\kappa^2\exp\left[2b-2(l+1)a\right],\\
c''-c'b'+m(c')^2+(l+1)a'c'-(m-1)\Lambda_{\rm (E)}\exp(2b-2c)
&=&-\frac{1}{2}\frac{l}{l+m}\kappa^2\exp\left[2b-2(l+1)a\right],\\
\frac{1}{2}m(m-1)(c')^2+\frac{1}{2}l(l+1)(a')^2+m(l+1)a'c'\\
\nonumber
-\frac{1}{2}m(m-1)\Lambda_{\rm (E)}\exp(2b-2c)
-\frac{1}{2}l(l+1)\Lambda_{\rm (L)}\exp(2b-2a)&~&=
\frac{1}{4}\kappa^2\exp\left[2b-2(l+1)a\right].
\end{eqnarray}

\setcounter{equation}{0}
\section{Dynamical system}

As is common practice for this type of system we may use Misner variables
\cite{misner} to rewrite the equations in a more familiar form.
By picking a suitable gauge these variables put the equations into
the form of the equation of motion 
for a particle in a potential.
To see this we define the following gauge choice,
\begin{eqnarray}
\label{gauge}
b&=&(l+1)a+mc,
\end{eqnarray}
and introduce a new variable,
\begin{eqnarray}
\label{misnervariable}
A&=&la+mc.
\end{eqnarray}
With these we find
\begin{eqnarray}
\label{basiceqns}
A''&=&l^2\Lambda_{\rm (L)}\exp(2A)
      +m(m-1)\Lambda_{\rm (E)}\exp\left[\frac{2(l+1)}{l}A-\frac{2(l+m)}{l}c\right],\\
\nonumber
c''&=&(m-1)\Lambda_{\rm (E)}\exp\left[\frac{2(l+1)}{l}A-\frac{2(m+l)}{l}c\right]
   -\frac{1}{2}\frac{l}{l+m}\kappa^2\exp(2mc).
\end{eqnarray}
We recognise these as describing the motion of a particle in the
$A-c$ plane, with a position dependent force acting on it.
It is noted that these equations may be derived from the following Lagrangian,
\begin{eqnarray}
{\cal L}&=&\frac{l+1}{l}(A')^2-\frac{m(l+m)}{l}(c')^2
+l(l+1)\Lambda_{\rm (L)}\exp(2A)\\
\nonumber
&~&+m(m-1)\Lambda_{\rm (E)}\exp\left[\frac{2(l+1)}{l}A-\frac{2(m+l)}{l}c\right]
+\frac{1}{2}\kappa^2\exp(2mc).
\end{eqnarray}
If we calculate from this the Hamiltonian, then one finds that the constraint
equation restricts us to zero energy solutions of this dynamical system.

For the explicit solution of these equations the following will be useful,
\begin{eqnarray}
\frac{\partial^2 y}{\partial \xi^2}&=&-\alpha^2\exp(2\beta y)\\
\Rightarrow\;\;y&=&-\frac{1}{\beta}
       \ln\left[\frac{\alpha}{c}\sqrt{\beta}\cosh[c(\xi-\xi_0)]\right],\\
\frac{\partial^2 y}{\partial \xi^2}&=&\alpha^2\exp(2\beta y)\\
\label{sinh_eqn}
\Rightarrow\;\;y&=&-\frac{1}{\beta}
       \ln\left[-\frac{\alpha}{c}\sqrt{\beta}\sinh[c(\xi-\xi_0)]\right],
\end{eqnarray}
where the range of $\xi$ for (\ref{sinh_eqn}) is 
\mbox{$\;\;-\infty<\xi<\xi_0$} and $c(>0)$, $\xi_0$ are arbitrary constants.

\setcounter{equation}{0}
\section{Melvin solutions}

We shall start our tour of solutions by showing that the above formalism
correctly reproduces what we already know, namely the Melvin solution.
This case has a Minkowski world volume so $\Lambda_{\rm (L)}=0$, one also
has \mbox{$\overline{{\rm ds}}_{\rm (E)}^2={\rm d}\phi^2$} so $m=1$ and because 
$\overline{{\rm ds}}_{\rm (E)}^2$ is one dimensional then $\Lambda_{\rm (E)}=0$. 
Taking $\Lambda_{\rm (L)}=0$, $\Lambda_{\rm (E)}=0$
but keeping $m$ general we find,
\begin{eqnarray}
\label{Aeqn}
A&=&c_0(\xi-\xi_0)\\
c&=&-\frac{1}{m}\ln\left[\frac{\kappa}{c_1}\sqrt{\frac{lm}{2(l+m)}}\cosh(c_1(\xi-\xi_1))\right]
\end{eqnarray}
the constraint equation requires $c_1=c_0\sqrt{\frac{m(l+1)}{l+m}}$, 
with $c_0(>0)$, $\xi_0$ and $\xi_1$ being
arbitrary constants. To make connection with the familiar form of the
Melvin solution we change coordinates to \mbox{$\xi=\ln(r)$}
and take $m=1$. Upon choosing
the constants \mbox{$\xi_0=0$} and 
\mbox{$\xi_1=\ln\left(\frac{2}{\kappa}\sqrt{\frac{2(l+1)}{l}}\right)$}
we find
\begin{eqnarray}
{\rm ds}^2 &=&{\rm ds}^2_{\rm (L)}\left[(\beta r)^{2c_0}+1\right]^{2/l}
              +{\rm d}r^2 r^{2c_0-2}\left[(\beta r)^{2c_0}+1\right]^{2/l}
              +{\rm d}\phi^2 r^{2c_0}/\left[(\beta r)^{2c_0}+1\right]^2
\end{eqnarray}
where \mbox{$\beta=\frac{\kappa}{2}\sqrt{\frac{l}{2(l+1)}}$}. 
We are then left with another constant, $c_0$, which we are free to choose.
The natural choice for $c_0$ is determined by regularity at the origin,
choosing $c_0=1$ means that as, $\rho\rightarrow 0$, 
the metric approaches Minkowski space and so is regular. Thus we have 
reproduced the Melvin solution (\ref{melvinmetric}).

It is instructive to look at the curvature invariants in the limits
\mbox{$\xi\rightarrow\pm\infty$} in this well known case, as we shall
be doing a similar analysis for the new solutions.
\begin{eqnarray}
c(\xi\rightarrow\infty)&\rightarrow& -\xi,\\
a(\xi\rightarrow\infty)&\rightarrow& \frac{2}{l}\xi,\\
b(\xi\rightarrow\infty)&\rightarrow& \frac{l+2}{l}\xi,\\
c(\xi\rightarrow-\infty)&\rightarrow& \xi-\exp(2[\xi-\xi_1])+{\rm const},\\
a(\xi\rightarrow-\infty)&\rightarrow& \frac{1}{l}\exp(2[\xi-\xi_1])+{\rm const},\\
b(\xi\rightarrow-\infty)&\rightarrow& \xi+\frac{1}{l}\exp(2[\xi-\xi_1])+{\rm const}.
\end{eqnarray}
We see then that in this orthonormal basis the components of the
curvature two forms vanish as \mbox{$\xi\rightarrow\infty$} and are
constant as \mbox{$\xi\rightarrow-\infty$}. As this is the orthonormal
frame then all invariants made from 
contractions of (\ref{curvature_two_forms}) will
be finite.
We shall see later that this is peculiar to the Melvin case, where
\mbox{$m=1$}, in general the space will be singular as 
we approach the fluxbrane core \mbox{$\xi\rightarrow-\infty$}.
For $m=1$ there are no such curvature components as $R_{{\rm (E)}ij}$
because ${\rm ds}^2_{\rm (E)}$ is one dimensional, so the divergence
of $(c')^2 \exp(-2b)$ in $R^i_{{\rm (E)}j}$ is not a problem.
For $m>1$ however these curvature
components are there and do diverge at the core of the fluxbrane leading
to a naked curvature singularity. We shall see however that at an
infinite proper distance from the core (for $\Lambda_{\rm (L)}\geq 0$)
the curvatures are well behaved, giving a regular asymptotic
spacetime.

\begin{figure}
\center
\epsfig{file=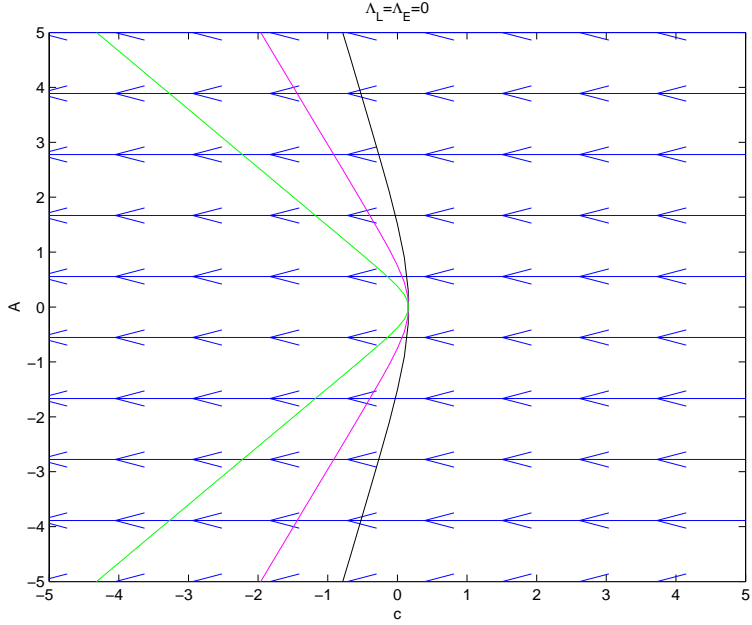,width=10cm}
\flushleft
\caption{Solutions for the Ricci flat world-volume and Euclidean manifold.
The core of the fluxbrane, $\xi\rightarrow -\infty$, is at
$A,c\rightarrow -\infty$.}
\label{fig:L0E0}
\end{figure}

As we have written out the equations in the form of a dynamical system
we may visualize this solution by considering the acceleration of
a particle with location in the $x-y$ plane of $(x,y)=(c,A)$.
Fig. \ref{fig:L0E0} is a plot with this in mind. Here the arrows show the
direction of the acceleration that the particle feels, 
given by $(c'',A'')$ of
(\ref{basiceqns}), but we have suppressed the magnitude of the
acceleration for clearer presentation.

The different trajectories are governed by the integration constants and
we show trajectories which only differ in the choice of $c_0$. 
From (\ref{Aeqn}) we see that $c_0$ characterizes the speed of the
particle in the $A$ direction, so a large value will tend to give a
straighter path in the $A-c$ plane.
The particle
starts at $c\rightarrow -\infty$ which represents the core of the
fluxbrane, moving to $c\rightarrow \infty$ which describes the region far
away from the fluxbrane.
In the case of $m=1$ we get the Melvin solution by choosing the $c_0=1$
trajectory, which leads to a regular core.

\setcounter{equation}{0}
\section{First extension, $\Lambda_{\rm (L)}\neq 0$, $\Lambda_{\rm (E)}=0$, $\rm m\geq1$}

\subsection{$\Lambda_{\rm (L)}>0$, $\Lambda_{\rm (E)}=0$, $m>1$}

The problem in getting a general solution to (\ref{basiceqns})
is the term involving $\Lambda_{\rm (E)}$ so, to get a feel
for the system we are trying to solve we start by setting
$\Lambda_{\rm (E)}=0$. 
This leads to an unconventional transverse
space as $\overline{{\rm ds}}^2_{\rm (E)}$ is usually taken to be
a round sphere metric.
This simplification allows us to get the general solution, in a later
section we shall look at the more interesting case where this
restriction is dropped.
In this case the equations (\ref{basiceqns}) may be solved to give
\begin{eqnarray}
A&=&-\ln\left[-\frac{l\sqrt{\Lambda_{\rm (L)}}}{c_0}\sinh(c_0(\xi-\xi_0))\right],\\
\label{csoln}
c&=&-\frac{1}{m}\ln\left[\frac{\kappa}{c_1}\sqrt{\frac{lm}{2(l+m)}}\cosh(c_1(\xi-\xi_1))\right],
\end{eqnarray}
and the coordinate range is \mbox{$-\infty<\xi<\xi_0$}, noting that $\xi=\xi_0$ is
at infinite proper distance from the core of the fluxbrane at $\xi\rightarrow-\infty$.
The constraint gives the relation, 
\begin{eqnarray}
\label{c0c1}
c_1=c_0\sqrt{\frac{m(l+1)}{l+m}}.
\end{eqnarray}
To study the solution at infinite proper distance we expand the solution around
$\xi=\xi_0$.
\begin{eqnarray}
c(\xi\rightarrow \xi_0)&\rightarrow& constant,\\
a(\xi\rightarrow \xi_0)&\rightarrow& \frac{1}{l}\ln(1/(\xi_0-\xi))+constant,\\
b(\xi\rightarrow \xi_0)&\rightarrow& \frac{l+1}{l}\ln(1/(\xi_0-\xi))+constant.
\end{eqnarray}
Which shows that the curvature components go to zero, (\ref{curvature_two_forms}).
In fact, we can see why this happens by using a more appropriate coordinate
system,
$R=1/(\xi_0-\xi)^{1/l}$. As $\xi\rightarrow \xi_0$ then
$R\rightarrow 0$ and the asymptotic metric becomes
\begin{eqnarray}
{\rm ds}^2 &\sim& \overline{{\rm ds}}^2_{\rm (E)}
              +l^2{\rm dR}^2
              +R^2\overline{{\rm ds}}^2_{\rm (L)}.
\end{eqnarray}
If we take $\overline{{\rm ds}}^2_{\rm (L)}$ to be the deSitter metric,
\begin{eqnarray}
\overline{{\rm ds}}^2_{\rm (L)}&=&-{\rm d}t^2+l_0^2\cosh^2(t/l_0){\rm d}\Omega_l^2,
\end{eqnarray}
then we see that asymptotically the metric, ds$^2$, is just the product of
$\overline{{\rm ds}}^2_{\rm (E)}$ and Minkowski space.

\begin{figure}
\center
\epsfig{file=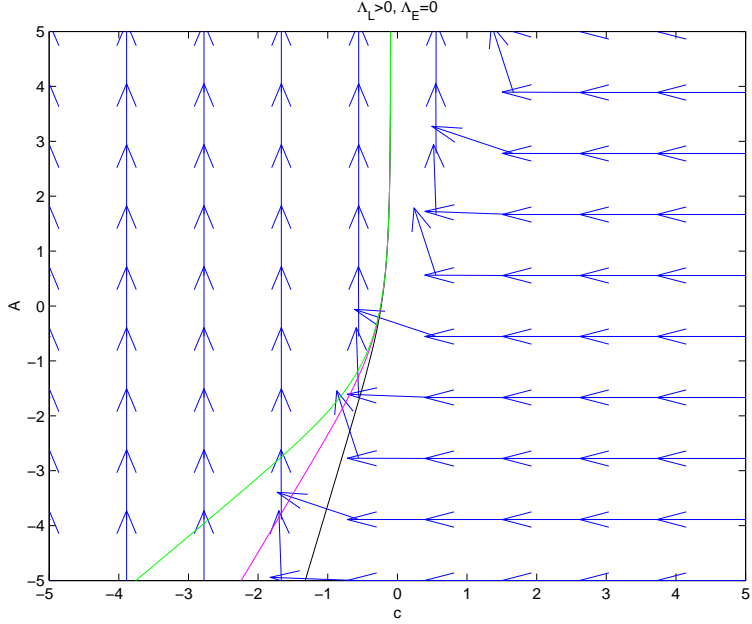,width=10cm}
\flushleft
\caption{Solutions for the positively curved 
world-volume and Ricci flat Euclidean manifold.
The core of the fluxbrane, $\xi\rightarrow -\infty$,
takes the values $A,c\rightarrow -\infty$.}
\label{fig:L1E0}
\end{figure}

The dynamical systems plot for this situation is given in Fig. \ref{fig:L1E0}.
Again the different trajectories are labelled by the integration constants,
here we show only what happens as $c_0$ varies.

\subsection{$\Lambda_{\rm (L)}<0$, $\Lambda_{\rm (E)}=0$, $m>1$}

Given the deSitter fluxbrane above then it is natural to look for
anti-deSitter fluxbranes that is, fluxbranes with negative curvature.
In order to get the general solution we again consider $\Lambda_{\rm (E)}=0$.
The starting point, (\ref{basiceqns}), now gives the solution,
\begin{eqnarray}
A&=&-\ln\left[\frac{l\sqrt{|\Lambda_{\rm (L)}|}}{c_0}\cosh(c_0(\xi-\xi_0))\right],\\
\end{eqnarray}
and $c$ is as before (\ref{csoln}), with the same relation holding between $c_0$ and $c_1$,
(\ref{c0c1}).
The coordinate range is $-\infty<\xi<\infty$ and we note that $\xi\rightarrow\infty$
is a finite proper distance from the core, so there is the possibility that we
are not covering the whole spacetime. However, by looking at limiting behaviour
of the scale factors,
\begin{eqnarray}
c(\xi\rightarrow\infty)&\rightarrow& -\frac{1}{m}c_1\xi,\\
a(\xi\rightarrow\infty)&\rightarrow& \frac{1}{l}(c_1-c_0)\xi,\\
b(\xi\rightarrow\infty)&\rightarrow& \frac{c_1-(l+1)c_0}{l}\xi,
\end{eqnarray}
we see that $b(\xi\rightarrow\infty)<0$ 
(because $(l+1)c_0>c_1$) so the scalar curvatures 
constructed from (\ref{curvature_two_forms})
diverge at a finite proper distance from the core. This leads to a rather
singular spacetime, with a singularity at a finite radial distance as
well as one at he core. This behaviour was seen previously \cite{gibbons88}
for the case of fluxbranes generated from a Maxwell 2-form, where the
worldvolume had a negative curvature.

\begin{figure}
\center
\epsfig{file=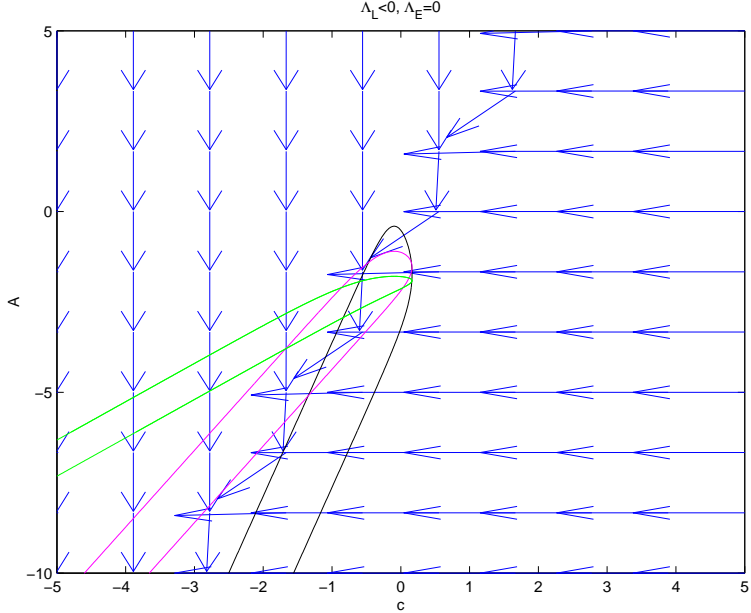,width=10cm}
\flushleft
\caption{Solutions for the negatively curved 
world-volume and Ricci flat Euclidean manifold.
The core of the fluxbrane, $\xi\rightarrow -\infty$,
takes the values $A,c\rightarrow -\infty$. The fact
that $A,c$ turn around and diverge back to
$-\infty$ is  the reason why there is a singularity
at finite proper distance from the core.
}
\label{fig:L-1E0}
\end{figure}

In this case the dynamical systems plot shows a marked difference with the
preceeding cases. Fig. \ref{fig:L-1E0} indicates some sample trajectories
that show how the particle doubles back on itself in the $c-A$ plane, it is
this behaviour that makes $b(\xi\rightarrow\infty)\rightarrow-\infty$
causing the curvature to blow up.

\setcounter{equation}{0}
\section{Second extension, $\Lambda_{\rm (L)}=0$, $\Lambda_{\rm (E)}>0$, $m\geq 1$}

The more interesting case is where $\overline{{\rm ds}}^2_{\rm (E)}$ is a round sphere
metric, rather than a Ricci flat one.
We can solve the equations if we make the ansatz that $A(\xi)$ is proportional
to $c(\xi)$, plus a constant. 
This means that we don't get the most general solution but, as we shall
argue, it is in fact the only sensible one.

For the two equations of motion to agree we find that
\begin{eqnarray}
A&=&\frac{lm+l+m}{l+1}c+\ln(\beta),\\
\beta^{(2l+2)/l}&=&\frac{\kappa^2}{2\Lambda_{\rm (E)}}\frac{lm+l+m}{(m-1)(l+m)}.
\end{eqnarray}
To find the solution for this we take the approach of substituting the
ansatz into the constraint equation, yielding
\begin{eqnarray}
c&=&-\frac{1}{m}\ln\left[-\kappa m\sqrt{\frac{l+1}{l+m}}(\xi-\xi_0)\right]
\end{eqnarray}
with coordinate range $-\infty<\xi<\xi_0$.
We note that this could, of course, have been derived using (\ref{basiceqns})
but the constraint equation would have meant taking a singular limit
of one of the constants of integration.
Again we find that the curvature two form components are finite as
$\xi\rightarrow \xi_0$, but blow up as the core of the fluxbrane is approached.
As an example we show the Ricci scalar,
\begin{eqnarray}
{\cal R}(\xi\rightarrow \xi_0)\rightarrow\left[\frac{2-3m-l}{ml}+\frac{1}{m^2l(l+1)}
                 +m(m-1)\Lambda_{\rm (E)}\right](\xi_0-\xi)^{2/m}\rightarrow 0
\end{eqnarray}
Although this isn't the most general solution it is the only one that
makes sense in the asymptotic region.
A perturbative analysis around this solution reveals that the perturbations
to the exact solution above take the form
\begin{eqnarray}
{\bf v}''&=&\frac{1}{(\xi-\xi_0)^2}{\rm M}(l,m){\bf v}
\end{eqnarray}
where \mbox{${\bf v}=(\delta A,\delta c)$} and M$(l,m)$ is some matrix.
It is found that the product of eigenvalues, ${\cal E}_1$ ${\cal E}_2$, of M$(l,m)$ is
\begin{eqnarray}
{\cal E}_1 {\cal E}_2&=&(-m^3-l^3-l^2m-4l^2m^2-2m^2l^3+m^2+2ml+l^2)
\end{eqnarray}
which is always negative for $l,m>1$. Thus we have one growing mode and
one oscillatory mode. However, because of the $1/(\xi-\xi_0)^2$ the oscillating
mode increases in frequency and amplitude causing the scale factors to
oscillate more and more wildly. For this reason we believe that the
particular solution found above is the relevant one.

\begin{figure}
\center
\epsfig{file=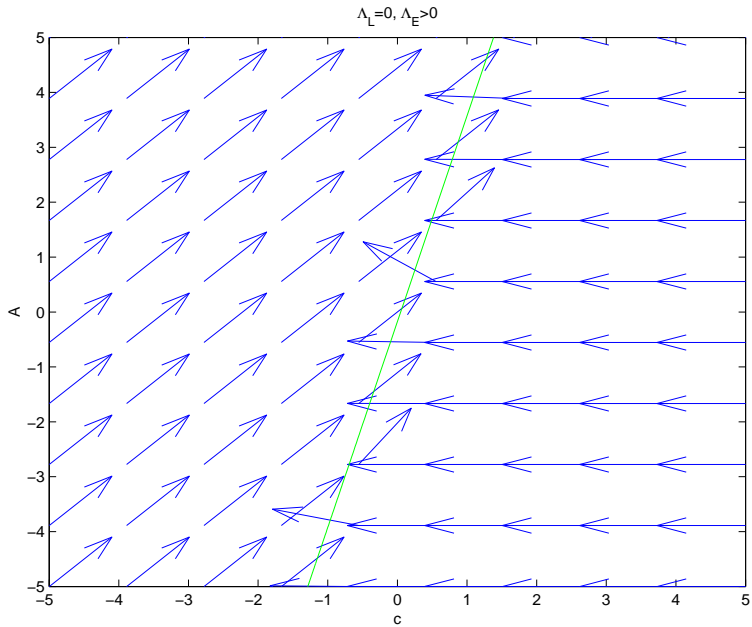,width=10cm}
\flushleft
\caption{Solutions for the Ricci flat
world-volume and positively curved Euclidean manifold,
eg. a round metric.
The core of the fluxbrane, $\xi\rightarrow -\infty$,
takes the values $A,c\rightarrow -\infty$.}
\label{fig:L0E1}
\end{figure}

We can get more intuition for this system by looking at the acceleration field
that the dynamical systems picture gives us, Fig. \ref{fig:L0E1}. If 
we place the particle
off the line marking the particular solution found above then it will accelerate
towards it. It is not an attractor however because of the magnitude 
of the acceleration, the particle oscillates about the particular solution with
an ever increasing amplitude and frequency.

\setcounter{equation}{0}
\section{Embedding diagram}

\begin{figure}
\center
\epsfig{file=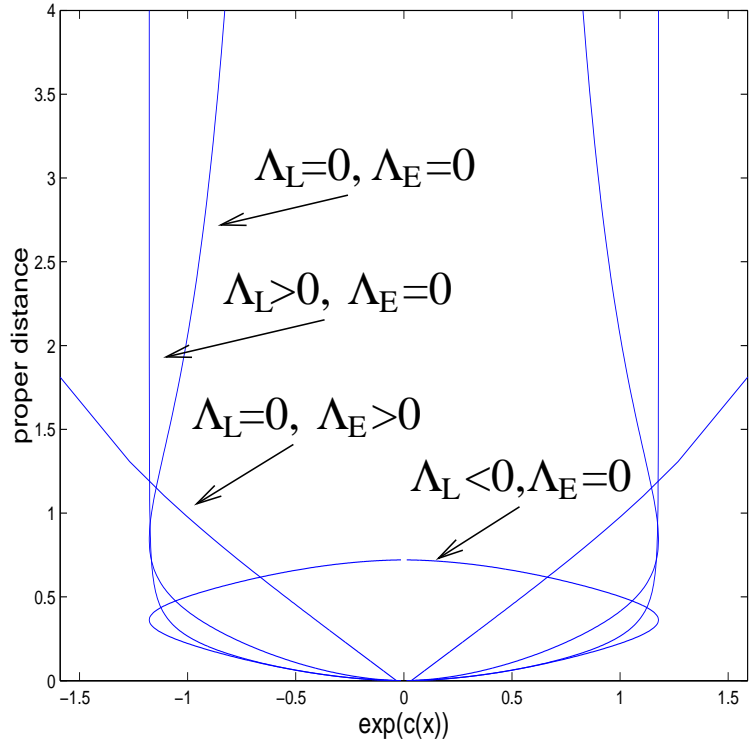,width=10cm}
\flushleft
\caption{
Here we show the embedding diagrams for the various solutions. They
describe how the volume of the Euclidean manifold, $\exp(c(\xi))$,
varies as we
move away from the core. To compare between cases we plot the volume
as a function of proper distance.
}
\label{fig:embed}
\end{figure}

The spacetime considered above can be neatly characterized by an embedding
diagram, showing how the size of the Euclidean manifold changes as a
function of proper distance from the core of the fluxbrane. 
Fig. \ref{fig:embed} shows such a diagram, where we recall from
(\ref{basic_metric}) that $\exp(c)$ characterizes the size of the Euclidean
manifold.
The cases where $\Lambda_{\rm (E)}=0$ are reminiscent of the 
embedding diagrams of \cite{gibbons88}, where the Euclidean manifold
was a circle and the form field was the Maxwell two form.

When plotted like this we see clearly that the AdS brane ($\Lambda<0$) is rather
singular, with it ending at a finite proper distance from the core.

\section{Dilatonic fluxbranes}

In supergravities one typically finds scalar fields which couple
to the field strength of the form field. It is therefore of
interest to see how these scalar fields modify the fluxbranes
described above. To study this we consider an action of the form,

\begin{eqnarray}
S&=&\int\left[R *1
              -\frac{1}{2}{\rm d}\phi\wedge *{\rm d}\phi
              -\frac{1}{2}\exp(\alpha\phi){\rm F}\wedge *{\rm F}\right].
\end{eqnarray}
So, we have a scalar field with an arbitrary dilaton type coupling in 
the Einstein frame. From this we may derive the following equations of motion,
\begin{eqnarray}
\label{formeqn}
{\rm d}\left[\exp(\alpha\phi)*{\rm F}\right]&=&0\\
{\rm d}*{\rm d}\phi&=&\frac{1}{2}\alpha\exp{\alpha\phi}{\rm F}\wedge *{\rm F}\\
R_{MN}&=&\frac{1}{2}e_M(\phi) e_N(\phi)
         +\frac{1}{2m!}\exp(\alpha\phi)\left[ {\rm F}_{M...}{\rm F}_N^{...}
                                             -\frac{m}{(m+1)(l+m)}{\rm F}^2 g_{MN}\right]
\end{eqnarray}
where the $e_M$ are the vectors dual to the one forms $e^M$.
For these solutions we take it that $\phi=\phi(\xi)$.
Using the same metric ansatz (\ref{basic_metric})
the form equation, (\ref{formeqn}), is solved for
\begin{eqnarray}
{\rm F}&=&\kappa\exp\left[-\alpha\phi-(l+1)a\right]
    e^\xi\wedge e^{l+2}\wedge e^{l+3}\wedge...\wedge e^{D-1},
\end{eqnarray}
with the Bianchi equation, d$*$F=0, being satisfied because 
$e^1_{\rm (E)}\wedge e^2_{\rm (E)}\wedge...\wedge e^m_{\rm (E)}$ is just the
volume form on $\overline{{\rm ds}}^2_{\rm (E)}$, up to a function of $\xi$.
Taking the same gauge choice as before (\ref{gauge}), 
and using the variable $A=la+mc$ (\ref{misnervariable})
we find the following equations of motion and constraint,
\begin{eqnarray}
\phi''&=&\frac{1}{2}\kappa^2\exp\left[2mc-\alpha\phi\right],\\
A''&=&l^2\Lambda_{\rm (L)}\exp(2A)
      +m(m-1)\Lambda_{\rm (E)}\exp\left[\frac{2(l+1)}{l}A-\frac{2(l+m)}{l}c\right],\\
c''&=&(m-1)\Lambda_{\rm (E)}\exp\left[\frac{2(l+1)}{l}A-\frac{2(m+l)}{l}c\right]
   -\frac{1}{2}\frac{l}{l+m}\kappa^2\exp(2mc-\alpha\phi),\\
0&=&\frac{l+1}{l}(A')^2-\frac{m}{l}(l+m)(c')^2-\frac{1}{2}(\phi')^2
-l(l+1)\Lambda_{\rm (L)}\exp(2A)\\
\nonumber
&~&-m(m-1)\Lambda_{\rm (E)}\exp\left[2\frac{l+1}{l}A-2\frac{l+m}{l}c\right]
-\frac{1}{2}\kappa^2\exp(2mc-\alpha\phi)
\end{eqnarray}

Due to the similarity with the dilaton free case we shall only consider 
$\Lambda_{\rm (L)}=0$.
Upon introducing new variables $Y$ and $Z$,
\begin{eqnarray}
Y&=&2mc-\alpha\phi,\\
Z&=&2\frac{l+1}{l}A-2\frac{l+m}{l}c,
\end{eqnarray}
we find,
\begin{eqnarray}
\phi''&=&\frac{1}{2}\kappa^2\exp(Y),\\
\label{Yeqn}
Y''&=&2m(m-1)\Lambda_{\rm (E)}\exp(Z)
      -\left[\frac{lm}{l+m}+\frac{1}{2}\alpha^2\right]\kappa^2\exp(Y),\\
\label{Zeqn}
Z''&=&2(m-1)^2\Lambda_{\rm (E)}\exp(Z)+\kappa^2\exp(Y).
\end{eqnarray}
Although we have been unable to find the general solution we are able,
just as before, to find a particular solution by taking the ansatz
$Z=Y+\ln\beta$. In order for (\ref{Yeqn},\ref{Zeqn}) to be consistent
we require,
\begin{eqnarray}
\beta&=&\frac{\kappa^2}{2(m-1)\Lambda_{\rm (E)}}\left[\frac{lm+l+m}{l+m}
                +\frac{1}{2}\alpha^2\right].
\end{eqnarray}
The solution then that satisfies the constraint equation is
\begin{eqnarray}
Y&=&-2\ln\left[-\frac{\zeta}{\sqrt{2}}(\xi-\xi_0)\right],\\
\phi&=&-\frac{\alpha\kappa^2}{\zeta^2}\ln\left[-c_1(\xi-\xi_0)\right],\\
\zeta^2&=&(m-1)\left[\frac{lm+2(l+m)}{l+m}+\frac{1}{2}\alpha^2\right]\kappa^2.
\end{eqnarray}
This solution has much the same properties as the analogous case without
the dilaton. The interpretation also follows from the previous case, with
us understanding the solution as the the fluxbrane which transmits the flux
between a dilatonic brane-antibrane pair of infinite separation.

\setcounter{equation}{0}
\section{Discussion}

In supergravity theories one typically requires antisymmetric form fields,
it has been the aim of this paper to construct the gravitational
analogue of a constant form field. The Melvin solution describes such
a scenario for a two-form (Einstein-Maxwell) where one finds that the
lines of magnetic flux self-gravitate into a cylindrical flux tube.
What we have found is that this is part of a more general picture, where
a magnetic $n$-form field strength in 
$D$ dimensional spacetime creates a fluxbrane
of $D-n-1$ spatial dimensions, a $(D-n-1)$-fluxbrane.
The physical origin of these objects is made clearer by thinking of
these fluxbranes as being formed between a 
$(D-n-2)$ brane anti-brane pair with infinite separation. 

In this paper we have performed a rather brute-force method of
finding solutions, however there is another, rather elegant,
way of getting the
dilaton-Einstein-Maxwell solution \cite{dowker95}.
The $D$ dimensional Melvin+dilaton solution comes 
from an unconventional
identification of points in $D+1$ dimensional Minkowski space.
When the dimensional reduction is then performed one finds that
the Kaluza-Klein vector is describing a magnetic fluxbrane.
The fluxbranes considered here are not expected to come from any
analogous technique as dimensional reduction of a metric does not
generate form fields, other than the Kaluza-Klein vector.
In this special case one is also able to generate the instantons
which describe the decay of this fluxbrane \cite{dowker95} by the
nucleation of charged spherical branes.
While we have not considered 
the stability of the generalized fluxbranes we
believe that, if classically stable, they could decay by similar
instanton processes. 

\section{Acknowledgements}
I gladly thank Dom Brecher, Ruth Gregory, Bert Janssen,
Simon Ross and Douglas Smith for useful conversations and
suggestions, and PPARC for financial support.


\end{document}